\documentstyle[twoside,fleqn,espcrc2,epsfig]{article}

\newcommand{\AmS}{{\protect\the\textfont2
  A\kern-.1667em\lower.5ex\hbox{M}\kern-.125emS}}

\title{The Rossi X-Ray Timing Explorer}

\author{J. H. Swank\address{Code 662, Goddard Space
Flight Center\\Greenbelt, MD 20771, U. S. A.}}

\begin{document}

\begin{abstract} RXTE has been operating for nearly 2 years and is
 planning the third. The spacecraft performance has been good and the
three
instruments are operating well. Observations have been made of the range
of targets suitable for RXTE, including such different objects as
accreting neutron stars and black holes, stellar flares, and supernova
remnants. The goals of studying high time resolution and  broad energy 
range and optimizing multi-wavelength participation are yielding
important  results. Oscillations found from low-mass X-ray binaries probably are 
signatures of the  spin of the neutron stars  and of the
shortest orbital periods around the neutron stars. These are constraining neutron
star parameters. Oscillation and
spectral results from black hole candidates bring into the realm of possibility
the possibilities of measuring the spins of the black holes and using X-ray data 
to test predictions of gravitation theory. Multi-wavelength observations are leading to
identification of the locations of the X-ray emission regions and, in the case of the
micro-quasars, to understanding of the
mechanisms for jet formation. 
Recently faster observing
response than originally planned has made possible 
some RXTE contributions to identification of
gamma-ray bursts.

\end{abstract}

\maketitle

\section{THE RXTE MISSION}

The RXTE was launched 1995 Dec 30. During the next week instruments were
turned on while the spacecraft was checked out and calibrated. The Guest
Observer (GO) program officially began Feb. 1, 1996, although a few time
critical observations were done during the in-orbit check out period for
the instruments. Some targets of opportunity, in particular, GRO J1744-28,
were observed. The transient, bursting pulsar had been discovered by BATSE
before RXTE's launch and interest in the unique object was high. Many
Target of Opportunity (TOO) proposals had been accepted for the first 9
months of the GO program, but remarkably, this object was not envisioned
and in fact there were no accepted proposals for a new, bright, pulsing
transient. The User's Group had ruled that requests for observation of
TOOs not covered by accepted proposals could be carried out
if the data were made public. Thus we started early on a substantial public
data base. Our planners make special effort to work new objects
into the schedule and to find when sources can best be
observed. In the two ensuing years 297 GO proposals were carried out,
71 of them (24 \%) as TOOs. The amount of good observing time was $3.52
\times 10^7$ s in 24 months, for a net  efficiency of 56 \%.

RXTE was planned to have features judged necessary to understand the
bright variable X-ray sky. It should have effective time resolution capable
of studying the dynamical time scales of neutron stars and stellar black
hole candidates, that is, better than tenths of milliseconds. For this,
the area should be large. In fact, the area flown was cut back to be
marginally effective for some goals. It should have energy coverage of at
least 2-200 keV. Galactic compact sources in the plane tend to have column
densities that cut off the observations at lower energies. Cyclotron lines,
synchrotron radiation, and inverse-Compton spectra require
sensitivity up to energies of  $\approx 100$ keV. 
Many spectra cut-off approaching 
$m_e c^2$, and it was impractical for us to achieve more than 200 keV
within our constraints. Fast and flexible response would allow
transient states to be studied and would facilitate multi-wavelength
observations; these  were also judged essential for this mission. 

\begin{figure}[t]
\parbox{73mm}{\epsfig{angle=0.,file=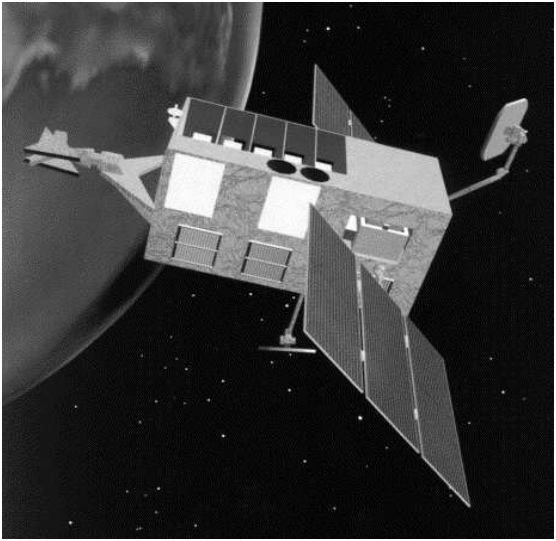,width=73mm,height=73mm}}
\caption{
Drawing of RXTE in orbit. The five PCA counters are in the
openings protected by solar shields. The 3 ASM Scanning Shadow
Cameras rotate on the boom on the left. The HEXTE clusters are hidden under the
thermal blankets to the right of the PCA. 
They look out the same direction as do the PCA
collimators, with the same field of view. 
Two oppositely directed antennas allow scheduled communication to
proceed regardless of changes in the observation schedule.
}
\end{figure}

In the launched configuration, the Proportional Counter Array (PCA) 
has an open area of
6250 cm$^2$ and the High Energy X-ray Timing Experiment (HEXTE) has
1600 cm$^2$ in phoswich scintillators. The Crab gives 13,400 counts
s$^{-1}$ in the PCA, 284 counts s$^{-1}$ in HEXTE. The time resolution of
the detector
electronics and the data system of the PCA is $1 \mu$s, for HEXTE,
$8\mu$s. There is a selection of data modes and steady telemetry rates of
50-256 kbps are regularly achieved, the lower value relatively steadily,
the higher for periods of about a half hour, with the supplement of high rate
dumps of the spacecraft solid state data recorder.  These can be added by request with
a days notice unless the Shuttle is flying. The achievements of high time
resolution that I discuss below, show that this capability is successfully
achieved. 

The responses and resolutions of the detectors are nominal, except that
because of an electronic failure,
for 1 of 8 HEXTE detectors the pulse heights are not available. HEXTE does
have a larger than expected
dead time due to large particle events,
which reduces the sensitivity. Activation of the PCA during
passage of the SAA has complicated calibration of the background.
Calibrations of background and response will still be improved, but are
now good to a couple of percent \cite{Jahoda96,Jahoda98,Roth98}. Papers
presented in this volume attest to the ability
to use the PCA and HEXTE together to measure cyclotron line
characteristics and the spectra of active galactic nuclei (AGN).

Sources can be viewed with both the PCA and HEXTE 
when the bore-sights are farther than 30
degrees from the sun. This constraint is more relaxed than for most
missions and has helped with the observations of transients and time
critical behavior. Of course even this constraint can be frustrating. When
GRO J1744-28 had a second outburst and for the second time BATSE saw that
the first day's bursts were of a different character than later,
with only  3 minutes separation time,
we were unable to make observations. Many monitoring observations
have followed sources for the duration of the mission. Brief looks (1000 s
minimum) are possible for campaigns of various durations and density.
Observations shorter than that, while sufficient to determine flux well,
turned out to stress the satellite slewing and attitude determination
resources. 

After the delivery of weekly observing plans, many re-plans are carried out
to allow for TOOs and to change the observing plans on the basis of
results. We had designed the system to the requirement that a new source
could be observed within 7 hours of discovery. Within the regular planning
process that is difficult, but such a rapid turn around is rarely optimal.
However the Beppo-SAX discovery of the X-ray
after-glows of Gamma-Ray Bursts has inspired development of much faster
procedures \cite{Marshall98}, which lead to observations within 2-3
hours. We can slew directly to a source (at 6 degrees per minute), 
if commanding is
available and there are no conflicting stored commands.

Figure 1 shows the satellite with the All Sky Monitor (ASM) on one end, comprising three 
one-dimensional
imaging proportional counters with coded masks that rotate around the boom
on which they are mounted and survey 70 \% of the sky. Solutions for the
flux of each cataloged
source are computed for each 100 s dwell in the step-wise rotation per satellite orbit of 90
minutes. Residuals to the fit indicate uncataloged sources.  

The five detectors of the Proportional Counter Array 
operated flawlessly for
about three months, at which time, the 
two detectors which operated at the coldest
temperatures, within an orbit of each other suffered short-lived breakdown
events. We have implemented the spacecraft's capacity to "look" at the data
being produced and take prescribed actions to turn high voltage off in less than a
minute if a breakdown event is
beginning. This, and operating the spacecraft slightly differently to
keep the detectors warmer, have so far allowed us to keep operating without
accelerated detector degradation.  During the first months the ASM
detectors lost 4 anode wires out of 24, due to high voltage breakdown. The
work-around in the ASM case was to burn the conducting carbon off the
affected anodes and to guard against operating at very high rates. Each
wire is calibrated \cite{Rem97a,Levine98} and the results for a large
number of sources can be
found
on the Web, on the MIT and the RXTE GOF pages: 
http://space.mit.edu/$\sim$derekfox/xte/ASM.html
http://heasarc.gsfc.nasa.gov/docs/xte.

RXTEs targets include galactic compact stars, AGN, stars, supernova
remnants, diffuse emission, and gamma-ray bursts. Galactic compact stars
account for 80 $\%$ of the targets: accreting high magnetic field binary
pulsars, accreting low-mass binary pulsars, rotation-powered pulsars,
black hole (BH) candidates, and white dwarfs. In this paper I will describe 
results that have been obtained for these sources. Other papers
in this volume describe some of the results on AGN and diffuse sources. 
References must
necessarily be incomplete
and
my choices of examples could well be biased
by my own associations.

\section{RXTE RESULTS ON GALACTIC SOURCES}

\subsection{Low-Mass X-ray Binaries}

The thermonuclear flash bursts have been convincing evidence that
the compact object in bursters, and by extrapolation, Z and atoll low-mass
X-ray binaries in general, is a neutron star. Yet it has remained uncertain
why no pulsed flux had been seen. Surface fields of $10^8-10^{10}$ G
fields are expected if the sources are progenitors of millisecond
radio pulsars with such fields. The beat frequency
model of the "Horizontal Branch" quasiperiodic oscillations of the Z
sources implied such fields (e.g. \cite{Ghosh92}). 
Many aspects still remain uncertain and the
beat frequency 
model is now in question. But the
kilohertz
oscillations that RXTE has discovered, in the persistent flux
and in some of the thermonuclear flash bursts, are signatures that
may identify the spin. These signatures, in a phase space not previously 
accessible, are a new
light on
the nature of all of the oscillations shown by these sources. It is
clear that the dimensions and time scales are characteristic of
the regions very close to the neutron star 
and that interpretation in terms of effects
of strong gravity, like a marginally stable orbit \cite{Zhang97,Kaaret97},
and possibly frame dragging \cite{Stella97} are quite possible. 
More than 15 sources exhibit the kilohertz oscillations
\cite{vanderklis98} and 6 sources
exhibit 300-600 Hz oscillations during bursts \cite{Stroh98}.

The characteristics of the oscillations and their interpretations
are discussed extensively in other papers in these proceedings.
The oscillations are sometimes either not present or difficult to discern,
even in sources in which large amplitudes have been observed. 
Thus failure to 
discover them in one observation does not imply they will not appear in 
another.
We believe the phenomena are characteristic of these
sources.

In studying the correlation of the kilohertz oscillations with luminosity
and source state, the transient bursters are useful. Outbursts of
X1608-522 and Aquila X-1 have occurred which were identified with the ASM.
Series of observations at different luminosity levels showed that the
identification of the mode and its correlation with luminosity are not
simple, but may involve different
states of accretion \cite{Zhangw98,Zhangs98}. 
There appear to be 2 ranges of luminosity in
which an
oscillation goes through the same frequency range.

\subsection{Rotation-Powered Pulsars}

The number of radio pulsars that are known to emit non-thermal pulsed
X-rays is small. Yet medium energy X-rays are diagnostic of
competing models of the generation of the pulses. RXTE started observing
the Crab pulsar during the in-orbit check-out phase, and continues to
monitor the pulsar for possible changes that may be associated with
changes at other wavelengths. RXTE has made extensive observations of
B1509-58 (150 ms) and also observed B1821-24 (3 ms) in order to check the
time assignments
of events and the timing software to a sub-millisecond accuracy
\cite{Rots98,Marsden98}. The
weight of the evidence is that absolute times are precise to 8 $\mu$s and
some interesting small differences between peak times in the X-ray and
radio are observed. Pulsars could have around them
UN-pulsed synchrotron nebulae, as the Vela pulsar may.
But the Vela pulsar has pulsed gamma-rays, pulsations recently
identified in OSSE data, and pulsed soft X-rays seen with ROSAT. 
It was  a challenge to see how the pulsed flux 
fell off between the OSSE and ROSAT energy
ranges. 
RXTE has seen pulsed flux in the 2-20
keV range and the spectrum connecting the softer and harder
X-rays suggests that the pulse is formed at the polar caps rather than the
outer gap \cite{Strickman98}.

\subsection{Soft Gamma-ray Repeaters}

While only one of the 3 known soft-gamma-ray repeaters, SGR 1806-20, 
has been reported active during
the first 2 years
of RXTE, RXTE
observations of it were spectacular \cite{Kou96}. 
Hundreds of very
short bursts were detected, most a few millisecond in duration, some so
bright as to be a hundred times the Eddington limit for a neutron star
source, with a broad distribution in sizes and separations \cite{Alaa98}. 
Low-level flux is seen
between the bursts. 
Conclusions have not been announced, except
that the spectra do not have 
any clear cyclotron feature \cite{Marsden98b}.

\subsection{Accreting Pulsars}

Many observations of accreting pulsars have been carried out, in which
pulse to pulse variations are evident in real time data. While some of
these results have been presented at meetings, few have yet been
published. 

However the observations of the bursting pulsar GRO J1744-28 show some
diagnostic effects. The pulses are nearly sinusoidal, except possibly
during the brightest bursts themselves, when the saturation of the
detectors by the high event rate \cite{Jahoda98b} distorts the
shape of the pulse. During
the outburst peaks, the pulsed flux outside of the bursts was about twice
as bright as the Crab, and if the source is as far as 7 kpc, about twice
the Eddington limit (for cosmic abundances and a 1.4 M$_\odot$
star)\cite{Giles96}. The
OSSE data showed that during the bursts the pulses came late
\cite{Strickman96}. The PCA data showed that {\it after} the bursts, the
pulse phase lagged behind and then gradually recovered between the bursts
\cite{Stark96}. The behaviors of the pulse delay and the flux in
recovering were similar, although the flux recovered more quickly. 
The bursts provide
a sharp mechanism, apparently, that distorts the field configuration.
Quasiperiodic oscillations in the pulsed flux were triggered by some
bursts \cite{Kommers97}.

As the persistent flux
subsided, the pulse period shortened  gradually until it reached a
plateau. Interpreted in terms of the transition between spin up and down,
the field should be $4 \times 10^{11}$ G \cite{Psaltis96,Stark98} and
reduction in the pulsed flux may correspond to the centrifugal barrier to
accretion \cite{Cui97a}. 

GRO J1744-28 is in the galactic center region, subject to confusion at the 
level of one milliCrab. Both RXTE and the WFC on BeppoSAX have seen that the 
region contains numerous little known burst sources. On the other hand, 
anomalous pulsars with periods near 7 s, only a few millicrab as well, 
are  better isolated. Although their
spectra are very soft and the PCA is less sensitive to such spectra than
to the hard spectra exhibited by the bursting pulsar, 
strong constraints on possible binary orbit parameters
are being obtained 
from upper limits to the
Doppler shifts (for 1E 2259+586, $\le$ 0.03 lsec \cite{Mereghetti98a,Mereghetti98b}). 
Other observations may make it difficult to fit any companion star into the orbit
without assuming the coincidence of a nearly pole-on view.

\subsection{Black Hole Candidates}

As the mass accretion rate through the disk onto accreting black holes increases, according to
theory (e.g. \cite{Chen96}), the hydrodynamics of the flow and the state of the
disk changes. For luminosities in the range of a few percent of Eddington, there is
perhaps a corona of high temperature electrons 
which Compton scatter soft photons to a spectrum that approximates a 
power law with photon index 1.6 
below about 30 keV.
For luminosities tens of percent of Eddington, 
there is evidence for existence of an optically
thick disk with $ kT \le 1 $ keV, in ultra-soft transients. 
It often coexists with a non-thermal component which looks
like a power-law (with photon index $\ge$ 2) in the medium energy X-ray regime. Cygnus X-1
and other BH candidates have also exhibited some clear temporal signatures, notably
excess aperiodic variance, power spectral densities, and time lags of higher relative to lower
energies. The timing behavior, more than spectral uniqueness, appeared to define a very
high state of accreting BHs, with accretion rate approaching the Eddington limit.
Nova Muscae and GX339-4 are the two candidates for this state identified by GRANAT and
{\it Ginga}. RXTE has observed all these states from BH candidates, although the
classic soft and  bright transient often associated with a BH, characterized by a fast rise and
a 30 day e-folding
time decline
has not yet been seen It is overdue, considering the 1 per year
average of 1987-1995.

Cyg X-1 has been observed many times, for proposals with different goals; 
observations have different strategies and observing modes. It went into a soft state
for $\approx$ 3 months, anticorrelated with the hard X-ray flux measured by BATSE
\cite{Zhangs97}. The power spectral density changed from band-limited white noise with 
hints of broad QPO features to a
power law; the lag times changed from 0.01--0.1 s to $\le 0.01$ s; the coherence between 
high (6.5-13.1 keV and above 13.1 keV) and the lower energies was near 1 except during the
transitions between the soft and hard states, when it was 0.6--0.9. 
\cite{Cui97b}. 
Coherence near 1 has been
taken as witness
that the same
scattering function could be applied to all flares or shots. The  long lag times in the low
state and the transition states  have
been taken to imply an extended corona (e.g. \cite{Hua98,Nowak98}), more extended than in the
soft state \cite{Cui97b}. So far, a
definitive report on the existence of millisecond bursts has not appeared,  although 
the shot distribution includes a
significant fraction of very short shots \cite{Focke98}, whose existence may imply a more
complex model than a central source of soft photons at the center of a large scattering
cloud.
Spectra have been modeled with
several different approaches to the inverse-Compton scattering off high energy electrons (e.g.
\cite{Dove98,Gierlinski98,Cui98a}).

The measurements of some known sources, in particular 1E1740-29 and GRS 1758-28
\cite{Smith97}, and some
new transients, of which GRS 1737-31 is a good example \cite{Cui97c}, show that both the
spectral
and temporal characteristics of Cyg X-1 are shared by other sources with luminosities about
$10^{37}$ ergs s$^{-1}$. 

Although no new very bright ($\ge$ 1 Crab) BH transient has been discovered (by the
ASM, GRANAT, the CGRO
BATSE, the BeppoSAX WFC, or any other mission), the two recurrent transients GRO J1655-40,
and GRS 1915+105 became bright in the first few months of the mission, and have had
important impact on the RXTE mission. The visibility of the  companion star to GRO J1655-40 
has given a
precise measurement of the mass of the compact object as 7 M$_{\odot}$, making it one of the
best
established BH candidates. The similarities of the two sources \cite{Rem97b}, called
micro-quasars
because of the radio jets, makes it probable they are both BHs. Further
similarities in the
X-ray spectral parameters  and in fast oscillations sometimes exhibited have suggested that
these two sources may differ from others in the spin of the
BHs \cite{Zhangs97b}. The association of
BH spin
and jets fits in well with the possibility that AGN jet sources are associated with
spinning BHs. For both these sources fast oscillations are sometimes observed
\cite{Rem97b,Morgan97} that
could be the epicyclic frequencies in the disks around spinning BHs (with dimensionless
angular momentum near 1) \cite{Nowak97}, or Lens-Thirring precession of the disks
\cite{Cui98}. RXTE, infrared, and radio observations of GRS1915+105 have led to a detailed
connection between disk instabilities and the injections into the jets
\cite{Belloni97,Fender97,Mirabel98,Eikenberry98}. The association between X-ray dips and the
appearance of ejected and cooling material has promise of being a break-through in understanding
both Galactic BH sources and Quasars. It explains why jets contain discrete blobs. 

The transient BH candidates that RXTE has observed in the first two years are listed
in Table 1. The values given for recurrence and decay times and peak fluxes (of outbursts
observed with RXTE) are only
approximate. 

\begin{table}[hbt]
\caption{Transient BH Candidates Seen with RXTE }
\label{tab:periods}
\begin{tabular}{lrrr}
\hline
Source          &Recur	    & Decay	&Peak Flux    		\\
		&Days	    & Days	&mCrab			\\
\hline
GRS 1915+105	& $300$     & $600$	&$3000$			\\
GRO J1655-40	& $300$     & $400$	&$3000$			\\
4U 1630-47	& $600$     & $150$	&$300$			\\
Cyg X-1		& $900$     & $85$	&$500$			\\
GRS 1739-278	& $ $       & $150$	&$700$			\\
GRS 1737-31	& $ $       & $30$	&$30$			\\
XTE J1755-324	& $ $	    & $30$	&$180$			\\	   
X1354-644	& $3600$    & $60$	&$50$			\\
\hline
\end{tabular}
\end{table}

\subsection{ASM Transients}

The ASM and the PCA in scans are both detecting transients $\ge 30-100$ mCrab, even in the
Galactic Center region. They complement each other in that ASM identification leads to a PCA
observation and the light curve of a source detected with the PCA can, after the fact, be
determined with the ASM, even when it did not identify the source. Outside the
Galactic Center, the ASM can detect sources like Mkn 501. 

The ASM is giving a very much more complete record than available before of the behavior of
transients, the low level in between outbursts, the small flares that fail to develop into
major outbursts. There are many Be star pulsars active at low levels with recurrence times
of months to years. Table 2 lists some that RXTE has seen. The ASM is also identifying much
more clearly long quasi periods of LMXB and pulsars \cite{Levine98}. At the level of 10
mCrab it is very sensitive to long time-scale periodic and quasi-periodic variations.  

\begin{table}[hbt]
\caption{Recurrent Be Star Transients Seen with RXTE }
\label{tab:periods}   
\begin{tabular}{lrrr}
\hline
Source                &Recur      & Decay     &Peak Flux              \\
                      &Days       & Days      &mCrab                  \\
\hline
GRO J1744-28          & $300$     & $200$     &$2000$                 \\
1RXS J1708-40 	      & $   $     & $60$      &$200$                 \\
X0115+63              & $111$     & $10$      &$60$                  \\
X1145-619             & $200$     & $20$      &$100$                  \\
GS 1843+009           & $ $       & $50$      &$26$                  \\
GS 2138+568	      & $ $       & $20$      &$50$                   \\
EXO 2030+375	      & $30$	  & $5$	      &$40$		     \\
A 1833-076            & $ $       & $20$      &$6$                  \\
GRO J2058+42          & $53 $     & $ $	      &$8$                   \\
4U 0726-260	      & $35 $     & $ $       &$3$			\\
\hline
\end{tabular}
\end{table}

\subsection{White Dwarfs and Stars}

The numbers of white dwarf sources, Dwarf Novae, Intermediate Polars, and Polars and of
non-compact stars that are
bright enough at hard X-rays for a collimated instrument like the PCA are smaller than the
numbers of neutron stars, but there were 58 proposals carried out in the first two years.
Some important results that
depended on either the large area or the ability to carry out long monitoring campaigns
have shown their power. The Intermediate Polar XY Ari is an eclipsing system and 
repeated coverage of the eclipse of the white dwarf was able to identify the source of the
X-ray emission as small radius shells at the magnetic poles \cite{Hellier97a}. Some of the
observations caught
the source going into outburst, and the changes in the pulsed fraction before the change in
the luminosity give a detailed insight into the magnetic connection between the disk and the
white dwarf \cite{Hellier97b}. A very different source is $\eta$ Carina, in which a set of
{\it Einstein Observatory}, ROSAT, BBXRT, and ASCA observations over the last decade had
identified a variable hard source with a thin-thermal spectrum as associated with the star
itself in the extended dust cloud. The X-ray monitoring campaign found a long build up of
X-ray flux to a drop coincident with UV and radio drops 
and consistent with the suggested 5 yr. 
period. The X-ray observations also found a still mysterious $\approx$ 85 day 
period \cite{Corcoran97}. 

\section{CONCLUSION}

RXTE has clearly been very successful for a variety of studies. The quantity and detail of the
data
make complicated systems like high magnetic field pulsars a challenge to analyze
and complexity
of the calibrations has daunted some observers initially. However, the results that have
appeared are
exciting and the good results from some observations of sources only a few mCrab in
strength are inspiring. We look forward to many more anticipated results, to
answering the deeper questions that observations have raised, and to observing more
examples of the rare phenomena that we sometimes encounter that greatly improve our insight into
high energy sources. Compared to the long run, the time these sources spend in some of the most
interesting states is short, and there are few enough of them that the combination
gives us
relatively rare examples. We hope to use the hard won resources of RXTE and Beppo-SAX to be able
to study as
many of
these as possible. 

I would like to acknowledge  the  many people 
who have contributed in many ways to planning, building,
launching, operating, and using RXTE.

\end{document}